\documentclass[intlimits,twoside,a4paper]{article}

\usepackage{amsmath,amssymb}
\usepackage{graphicx}

\usepackage[T2A]{fontenc}
\usepackage[cp1251]{inputenc}

\usepackage{cmpj3}
\issue{2020}{23}{2}{23607}
\doinumber{10.5488/CMP.23.23607}

\title[Molecular--atomic fluid transformation in hydrogen]%
{Velocity autocorrelations across the molecular--atomic fluid transformation
in hydrogen under pressure
}
\author[G. Ruocco, T. Bryk, C. Pierleoni, A.P. Seitsonen]%
{G. Ruocco\refaddr{label1,label2}, T. Bryk\refaddr{label3,label4}, 
 C. Pierleoni\refaddr{label5,label6}, A.P. Seitsonen\refaddr{label7}}
\addresses{
\addr{label1} 
 Center for Life Nano Science @Sapienza, Istituto Italiano di
  Tecnologia, \\295 Viale Regina Elena, I-00161, Roma, Italy
\addr{label2} 
 Dipartimento di Fisica, Universita di Roma La Sapienza, I-00185, Roma, Italy
\addr{label3} 
Institute for Condensed Matter Physics of the National Academy of Sciences of Ukraine,\\
1 Svientsitskii St., 79011 Lviv, Ukraine
\addr{label4} 
Lviv Polytechnic National University, 79013 Lviv, Ukraine
\addr{label5} 
Universit\'e Paris-Saclay, UVSQ, CNRS, CEA, Maison de la Simulation, F-91191, Gif-sur-Yvette, France
\addr{label6} 
 Department of Physical and Chemical Sciences, University of L'Aquila
and CNISM UdR L'Aquila, \\Via Vetoio 10, I-67010 L'Aquila, Italy
\addr{label7} 
 D\'{e}partement de Chimie, \'{E}cole Normale Sup\'{e}rieure, 24 rue Lhomond, F-75005 Paris, France
}

\date{Received March 10, 2020, in final form March 31, 2020}

\begin{document}
\maketitle
\begin{abstract}
Non-monotonous changes in velocity autocorrelations across the transformation from molecular 
to atomic fluid in hydrogen under pressure are studied by \textit{ab initio} molecular dynamics 
simulations at the temperature 2500~K. We report diffusion coefficients in a wide range of 
densities  from purely molecular fluid up to metallic atomic fluid phase.
An analysis of contributions to the velocity autocorrelation functions from the motion of molecular 
centers-of-mass, rotational and intramolecular vibrational modes is performed, and 
a crossover in the vibrational density of intramolecular modes across the transition is discussed.


\end{abstract}

\section {Introduction}

Exotic properties of condensed matter at high pressures like a transition to non-metallic state in 
 Li~\cite{Mat09} and Na crystals \cite{Ma09} or emergence of ring-like jump diffusion of atoms in 
solid iron at conditions of the Earth's inner core \cite{Bel17,Bel19} caused huge interest in 
simulations and theory of matter at extreme conditions. Liquids at high pressures have their 
own specific features defined by topological disorder and partial localization of electrons 
\cite{Rat07,Tam08}.  
Especially in the field of dynamic properties of liquids at high pressures, there is a lack
of understanding the behaviour of collective modes as well as of single-particle dynamics, that 
is connected with possible transformations in atomic structure at high pressures \cite{Bry13}.
The situation with understanding the dynamic properties in molecular liquids which under
high pressure undergo dissociation of molecules and even transformation to purely atomic fluid 
is even more fascinating and less studied. According to the theory of collective dynamics \cite{Ber} 
such systems should be considered at least as a mixture of molecular and atomic components. One of 
the most simple and mysterious (because of the long-time search for its metallic state) molecular 
systems, which transform under pressure into atomic ones, are 
solid and liquid hydrogen \cite{McM12}.

Recently we observed in {\it ab initio} simulations of hydrogen fluid an interesting behaviour
of the long-wavelength charge fluctuations, which made evidence, that right in the transition
region from purely molecular to atomic hydrogen fluid there can appear unscreened ions with 
long-range Coulomb interaction \cite{Bry20}. This is in line with the recent results 
obtained with Coupled Electron-ion
Monte Carlo simulations focused on detection of the possible formation of stable 
molecular ions at the liquid-liquid phase transition in hydrogen \cite{Pie18}.
Hence, the molecular--atomic fluid transformation region can be much more sophisticated than 
just simply a mixture of neutral molecules and atoms, that must have some fingerprints in 
the dynamic properties of these fluids. 

Actually, the dynamic properties of hydrogen fluid at 
high pressures were not studied enough. Mainly the old experimental \cite{Duf94} and simulation 
\cite{Ala95} studies of the sound propagation in hydrogen fluid are well known. 
Having in mind the study of collective dynamics, which
is extremely sensitive to the existence of correlations between different species as well as to the
presence of ions with long-range Coulomb interaction \cite{Mar,Bry04}, we will first study  the 
single-particle dynamics across the transition region. In particular, we aimed at the study of 
velocity autocorrelation functions for hydrogen fluids in a wide range of densities: from molecular
liquid up to the metallic atomic liquid.  
The remaining paper is organized as follows: in the next section we report the details of 
our {\it ab initio} simulations, then will report our results for the velocity autocorrelation 
functions, their Fourier-spectra and diffusion coefficients, and the conclusion of this study 
will be given in the last section.

\section {Ab initio molecular dynamics simulations}\label{section:AIMD}

First-principles simulations for seven densities of hydrogen fluid were performed using 
a model system of 1000 particles at $T=2500$~K by the VASP package. The chosen temperature is 
high enough for application of classical equation of motions to protons. The starting 
configuration for the smallest density was taken from the classical simulations of molecular 
dimers. The standard $NVT$ ensamble was used for each density. Upon an equilibration of up to 
8 000 timesteps, we checked out the temperature and pressure fluctuations and then switched to 
production runs, each of 20 000 timesteps. The time step in {\it ab initio} simulations was
0.2~fs. The range of simulated densities corresponded to the pressures from 2.5~GPa (purely 
molecular fluid) up to 166.5~GPa (metallic atomic hydrogen fluid \cite{Bry20}).

The electron-ion interaction was the projector-augmented wave (PAW) potential 
\cite{Blo94,Kre99}, and the plane-wave expansion cut-off was 250~eV. The generalized gradient
approximation in PBE version \cite{Per96} was used in account for exchange-correlation effects.
In the sampling of the Brillouin zone we used only the $\Gamma$ point because of the large size 
of our simulated systems.

\section{Results and discussion}

\begin{figure}[!t]
\centerline{
\includegraphics[width=0.47\textwidth]{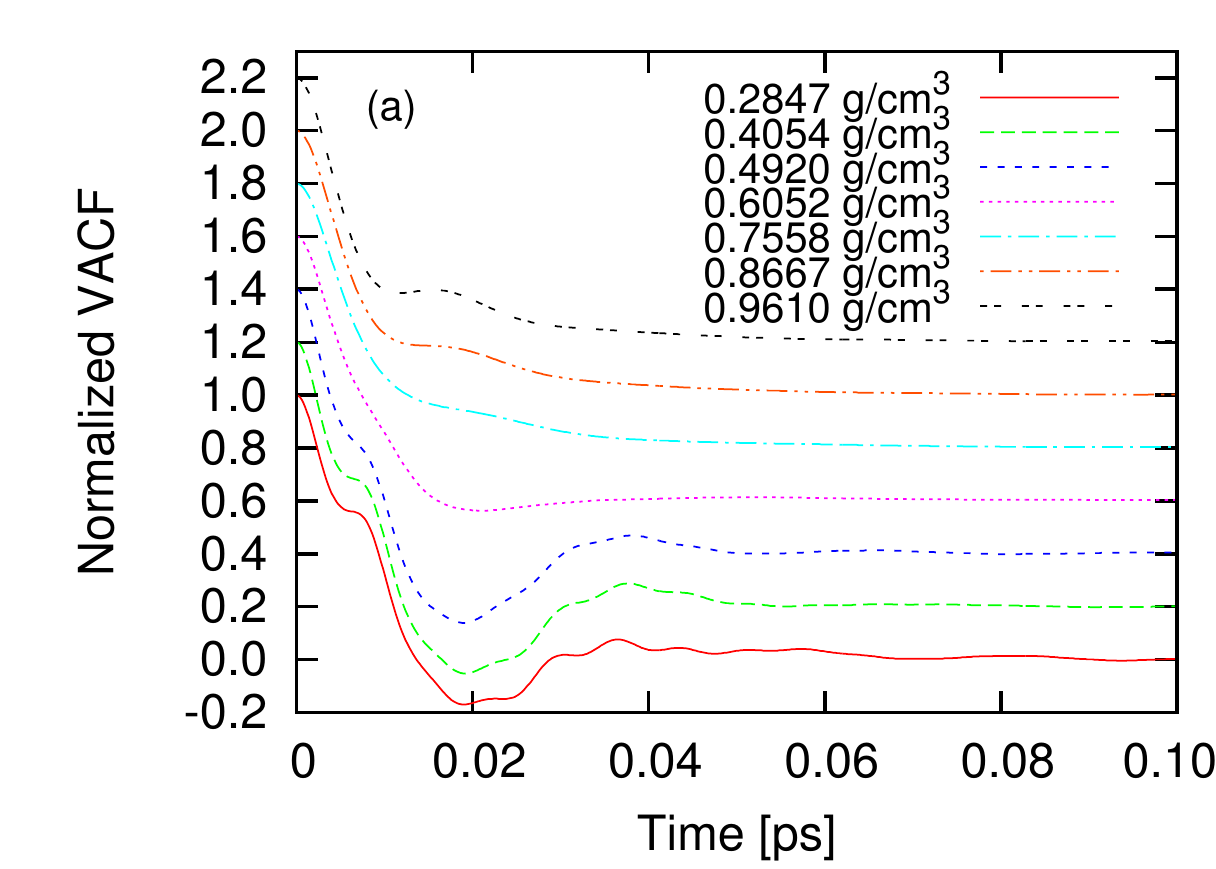}%
\includegraphics[width=0.47\textwidth]{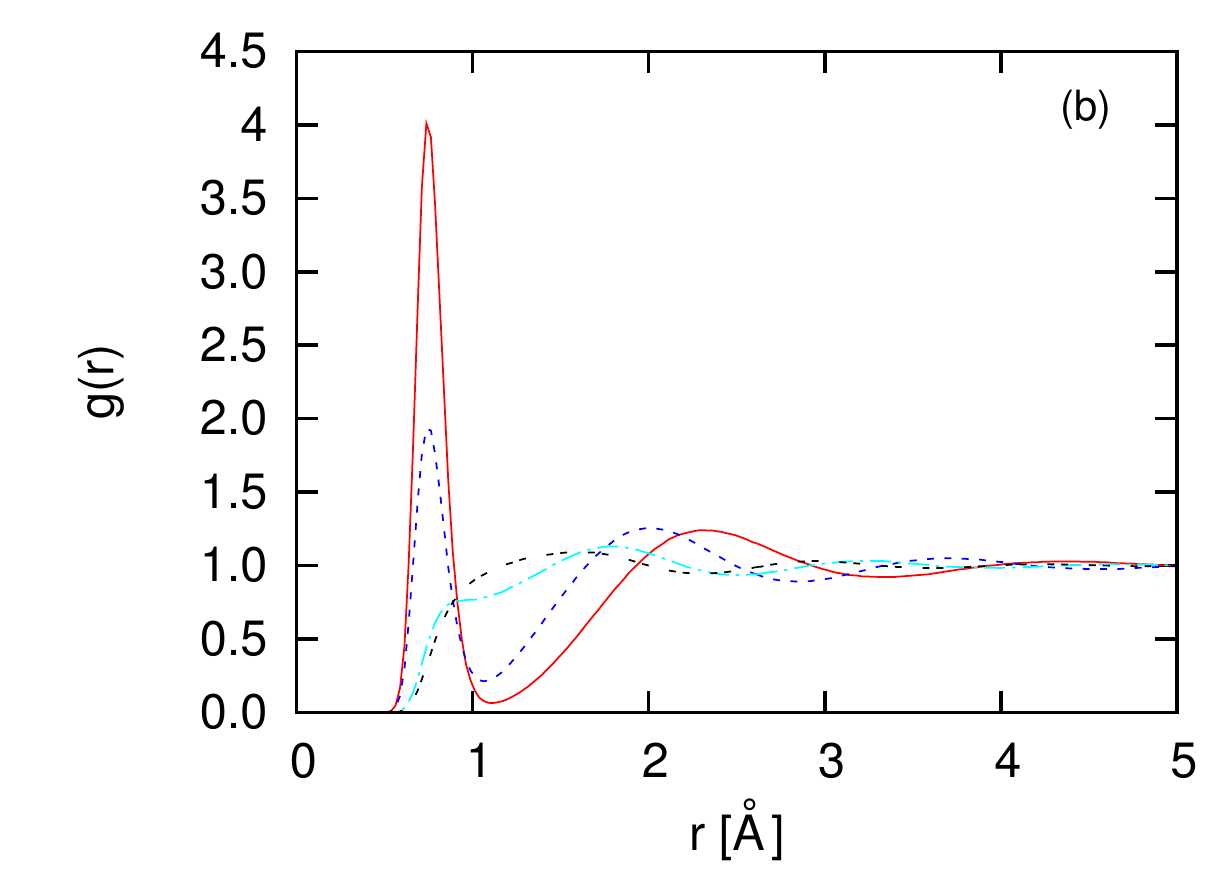}%
}
\caption{(Colour online) Non-monotonous density dependence of velocity autocorrelation functions (VACFs) 
in hydrogen fluid at $T=2500$~K (a) and pair distribution functions for four densities (b). 
For convenience, the VACFs in (a) were shifted vertically with the step 0.2. Different lines 
in (b) correspond to the same densities as in (a).
} \label{vacf_t}
\end{figure}

The main features in the behaviour of the structural and dynamic properties of hydrogen fluid in
a wide range of pressures are defined by the region of transformation from purely molecular 
to purely atomic fluid, in which the hydrogen molecules break up due to the pressure increase with 
simultaneous increase of a contribution from purely atomic subsystem. In figure~\ref{vacf_t}~(a) we show the 
density dependence of the velocity autocorrelation functions (VACF)
\[
\psi(t)=\langle{\bf v}_i(t){\bf v}_i(0)\rangle_{i,t_0}\,,
\]
which estimates the time correlation in proton velocity along the trajectory, 
averaged over all the
particles and over all the possible origins of the calculated time correlations $t_0$. In order to 
analyze the changes in VACFs, we plotted in figure~\ref{vacf_t}~(b) the pair distribution functions $g(r)$ for four 
densities. They make evidence of the pure molecular fluid at the lowest density 0.2847~g/cm$^3$ 
with the intramolecular peak of $g(r)$ centered at 0.75~\AA, while at the highest studied density
of 0.9610~g/cm$^3$, the $g(r)$ looks typical as for purely low-density atomic fluids. For 
intermediate densities, one observes a gradual reduction of the intramolecular peak and a 
transformation (in amplitude and with a shift towards the smaller distances) of the intermolecular 
peak into the first interatomic peak (at high densities). Here, we should  mention that although 
 the classical equations of motion of protons were considered, much more precise pair distribution 
functions and in general a picture of the structural 
features across the transformation from 
molecular to atomic hydrogen fluid are observed in the Quantum Monte Carlo simulations \cite{McM12}.

The typical time dependence of VACFs for dense fluids lies in the existence of the so-called cage-effect,
which is reflected by the negative region of $\psi(t)$ i.e., when the direction of particle velocity
at time $t$ is opposite (or rather almost opposite in order to result in the negative scalar product) 
with the initial velocity ${\bf v}_i(0)$ due to back-scattering on the nearest neighbours. 
In a wide range of densities, we do observe in figure~\ref{vacf_t}~(a) the cage-effect.
As a consequence of the molecular structure, we observed a combination of the high-frequency 
oscillations and the cage effect in the 
velocity autocorrelations, which are very pronounced, especially for the two lowest studied densities
[see figure~\ref{vacf_t}~(a)]. For density 0.4920 g/cm$^3$, the high-frequency oscillations are smeared out, while for 
the higher densities they 
are not observed. For the densities up to 0.6052 g/cm$^3$, the back-scattering effect is 
clearly observed, though for the higher densities studied here  the negative region of $\psi(t)$
vanishes and the VACFs show the time decay typical of gases. 
For the highest studied density 0.9610 g/cm$^3$, the VACF again becomes non-monotonously decaying 
in time showing a weak oscillation at time $\sim0.015$~ps, which  with the further increase 
of density
(and pressure) would perhaps increase in amplitude and should definitely again lead to the back-scattering 
effect at ultra-high pressures.

\begin{figure}[!t]
\includegraphics[width=0.45\textwidth]{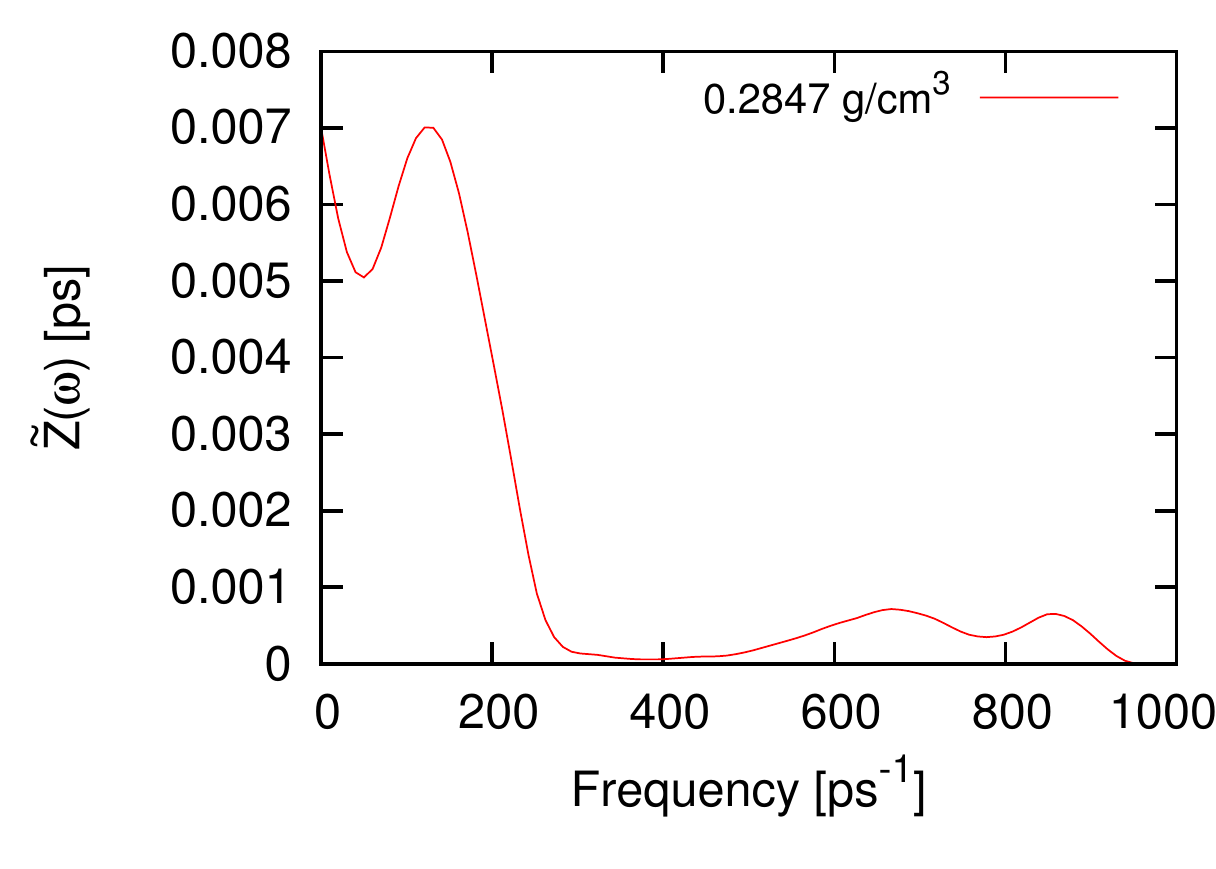}%
\includegraphics[width=0.45\textwidth]{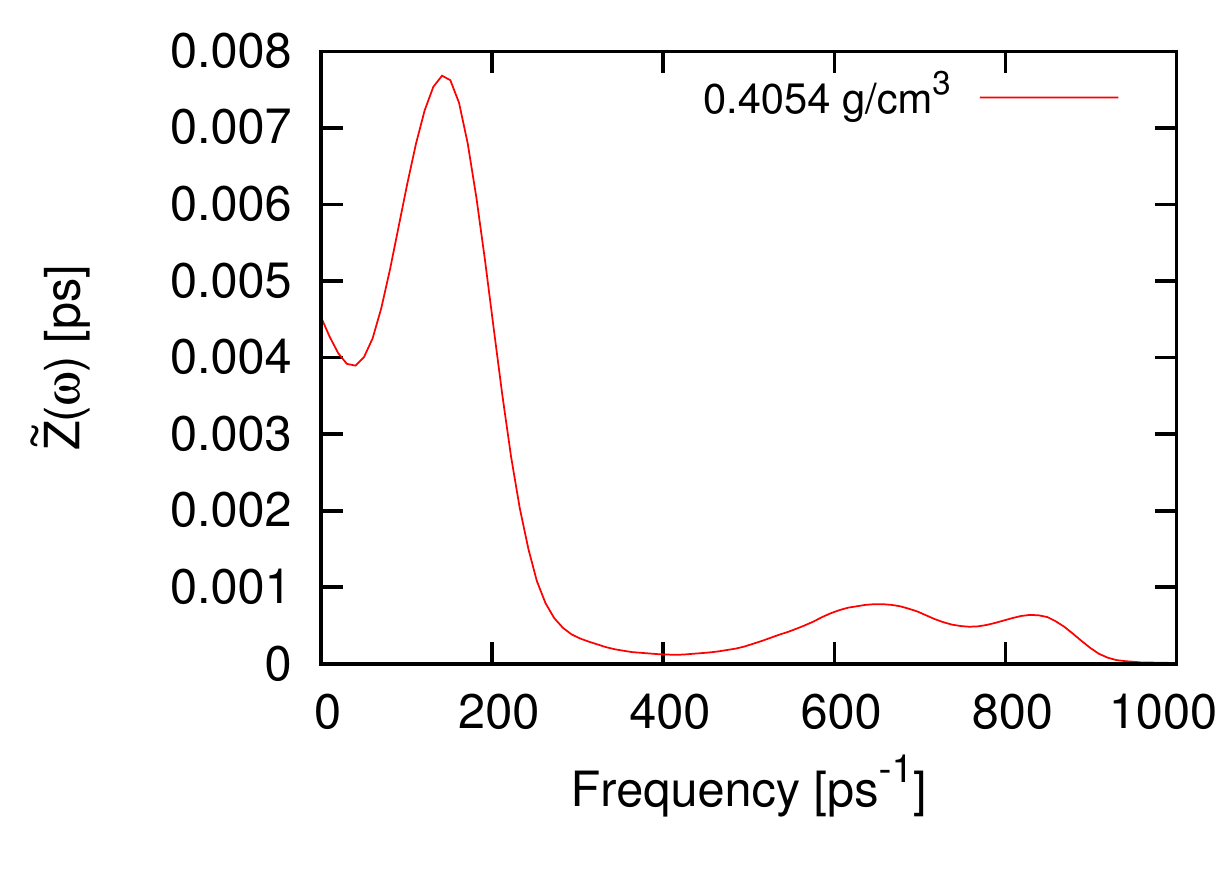}%

\includegraphics[width=0.45\textwidth]{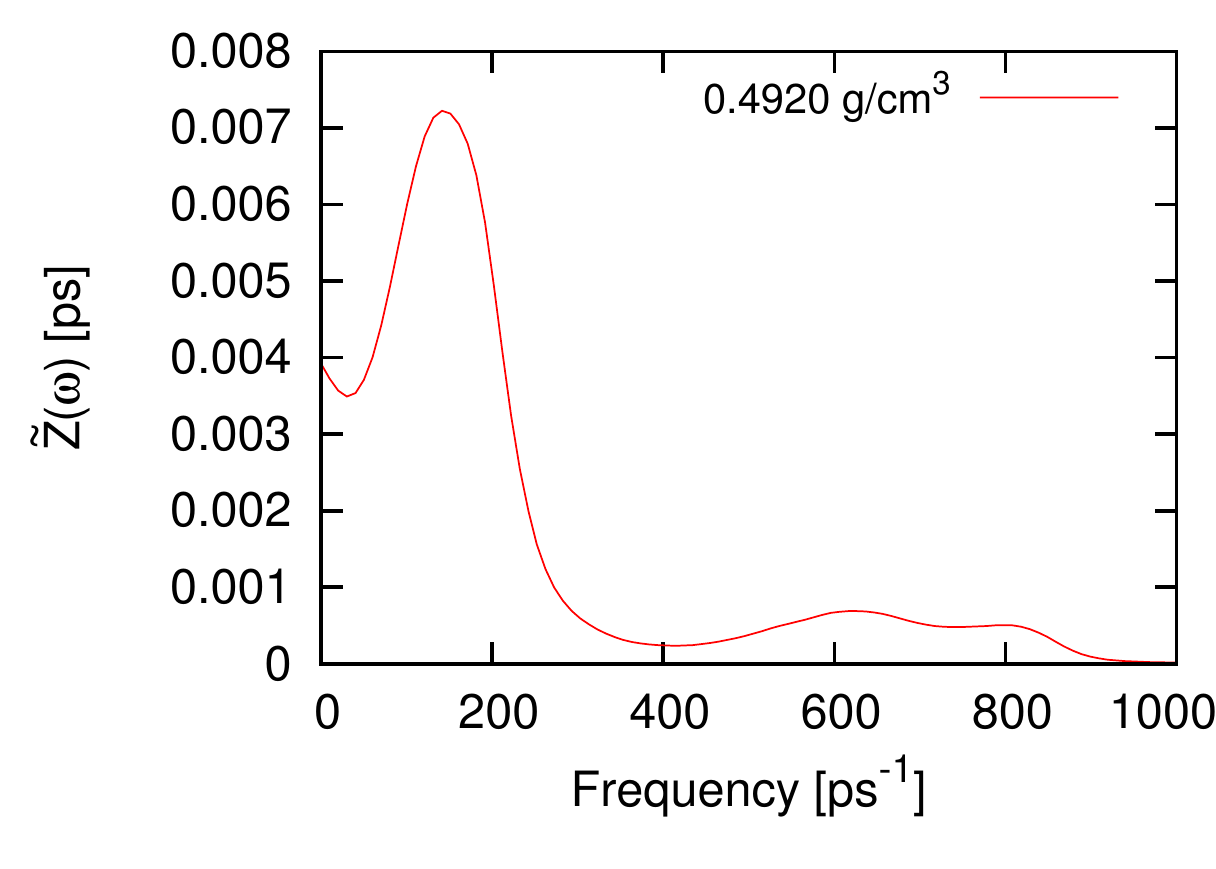}%
\includegraphics[width=0.45\textwidth]{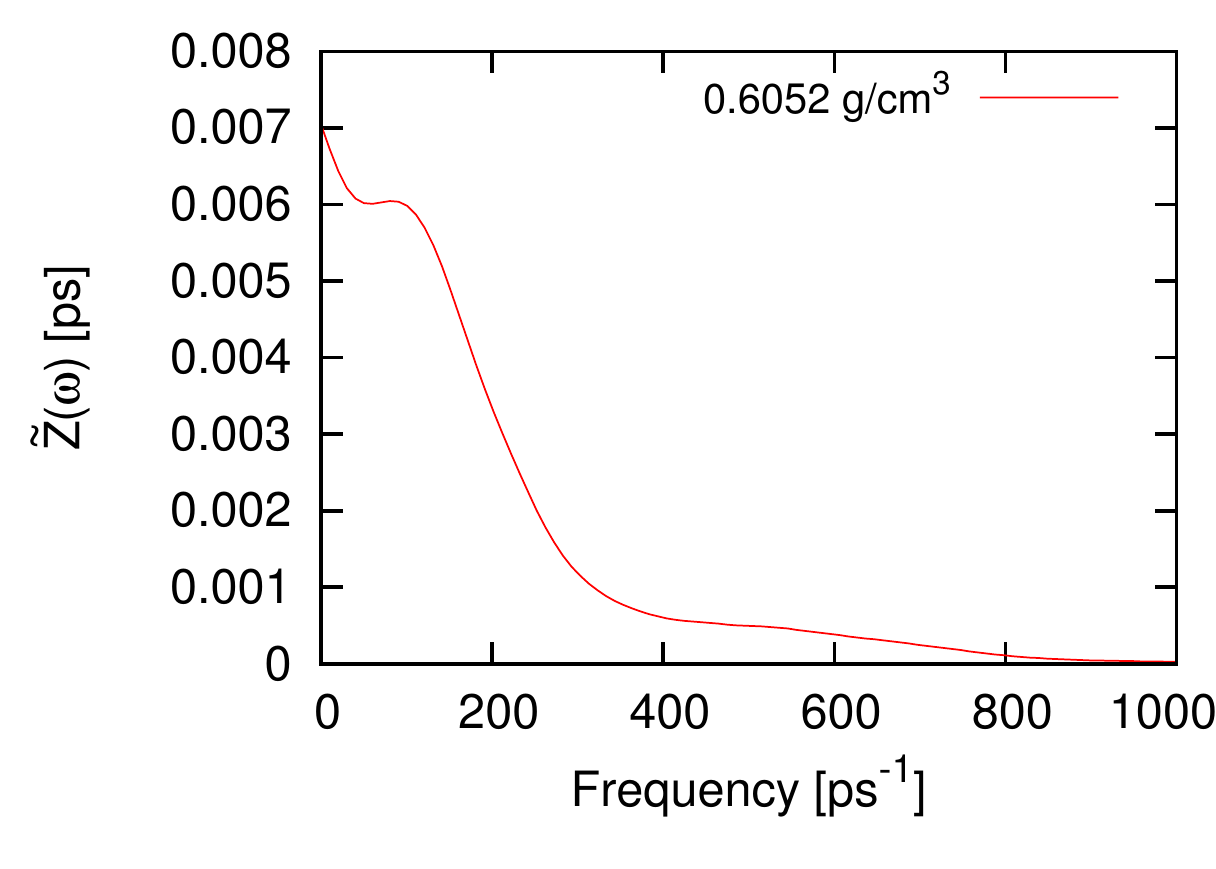}%

\includegraphics[width=0.45\textwidth]{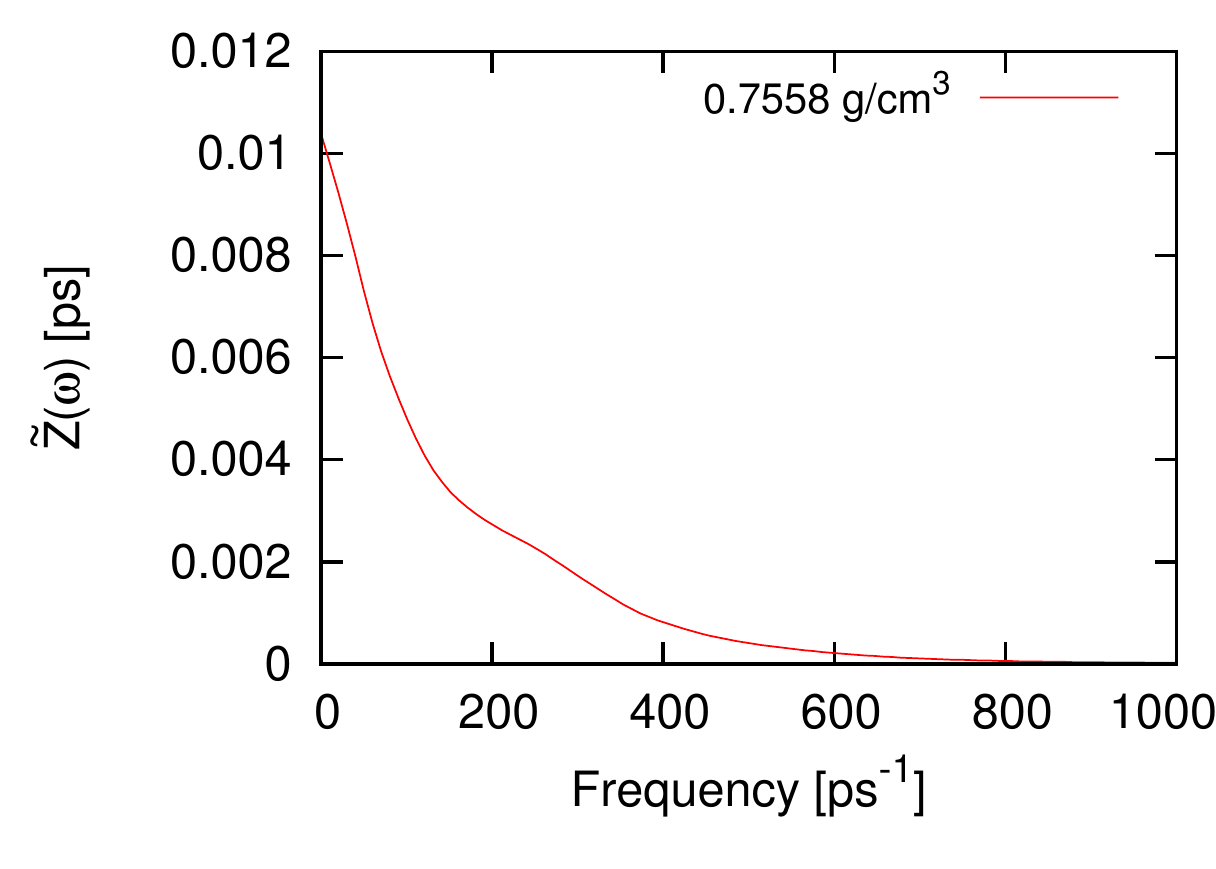}%
\includegraphics[width=0.45\textwidth]{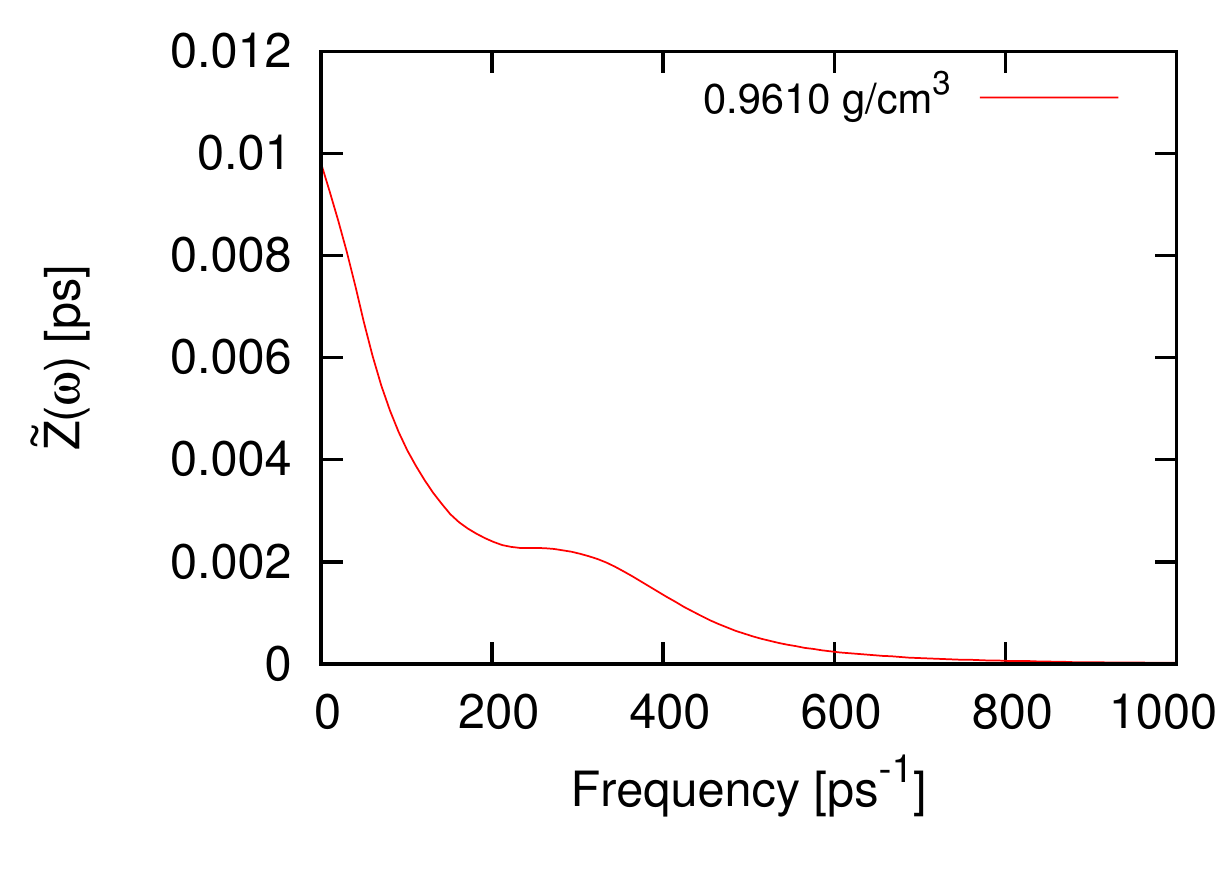}%
\caption{(Colour online) Frequency spectrum of normalized velocity autocorrelation functions for hydrogen 
at $T=2500$~K and different densities.
} \label{vacf_w}
\end{figure}

We calculated the Fourier-spectra of VACFs (figure~\ref{vacf_w}) in order to look how the low-frequency (mainly 
acoustic excitations and diffusion) processes and high-frequency modes change across the 
transformation from molecular to atomic hydrogen fluid. For the low densities up to 0.4920 g/cm$^3$,
the high-frequency intramolecular modes are well-defined in the spectra. Note that due to the 
rotational-vibrational coupling, one observes a splitting, i.e., two peaks in the region of frequencies
600--900 ps$^{-1}$. For higher densities, the Fourier-spectra of VACFs look monotonously decaying 
functions in the high-frequency region, and the well-defined peak in the region of frequencies 
$\sim180$~ps$^{-1}$ disappeares too. Herein below we will try to decompose the VACFs into contributions 
from the motion of center-of-mass, rotations and intramolecular modes in order to assign the 
different frequency regions to these processes. However,  we previously note that the Fourier-spectrum
at zero frequency, which actually is directly connected to the diffusivity via the Kubo 
integral, changes non-monotonously with density.

\begin{figure}[!t]
\centerline{
\includegraphics[width=0.6\textwidth]{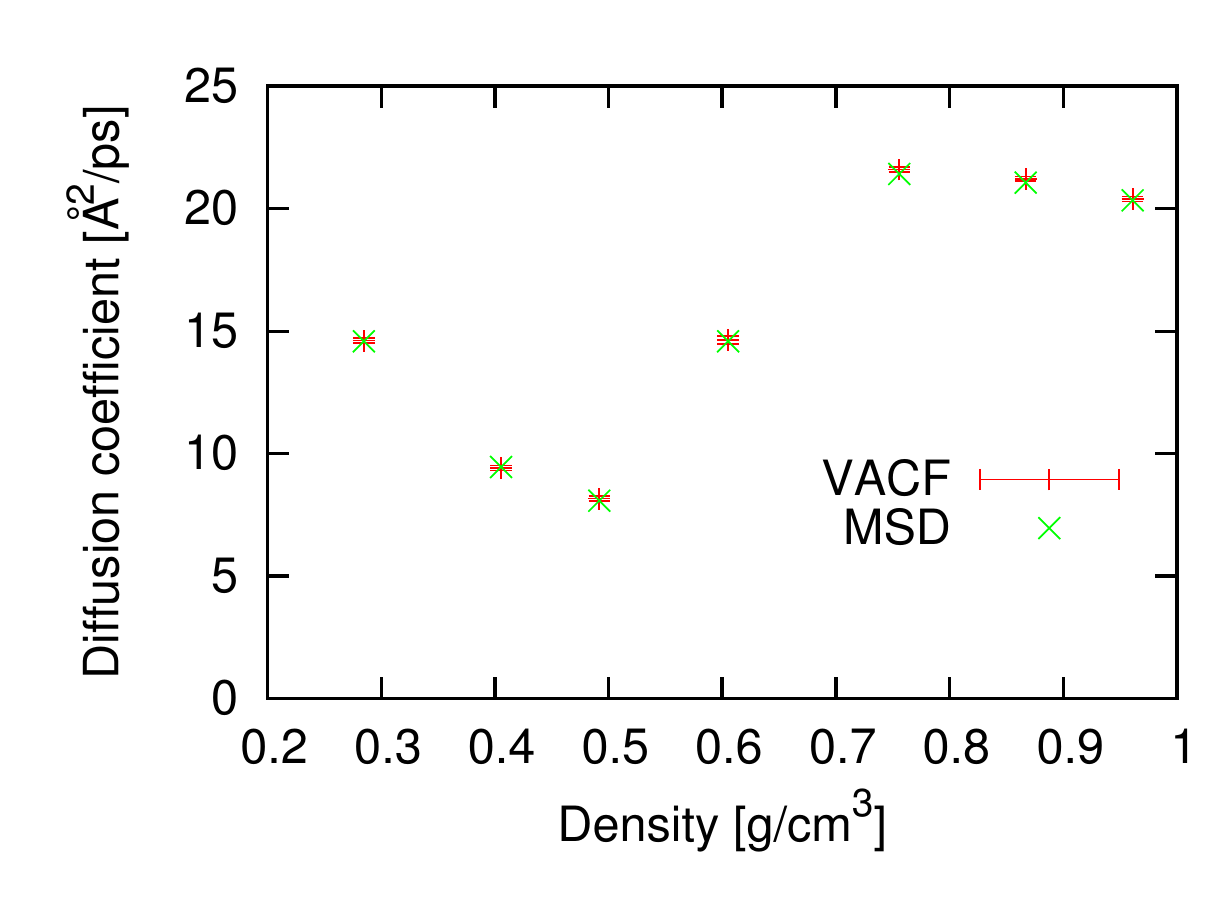}%
}
\caption{(Colour online) Non-monotonous density dependence of diffusion coefficients for hydrogen fluid 
at $T=2500$~K and different densities, 
calculated from the velocity autocorrelation functions (shown in figure~\ref{vacf_t}) by 
Kubo integral (plus symbols with error bars). A check of the diffusion coefficients via 
long-time asymptotes of the mean square displacements (cross symbols) shows very similar results.
} \label{diff}
\end{figure}

In figure~\ref{diff}, we show the density dependence of the diffusion coefficient $D$ calculated via Kubo integral
(plus symbols with error bars), and its check via calculations of the long-time asymptotes of 
mean-square displacements (MSD)
$\langle R^2\rangle(t)$ \cite{Han,Boo} (cross symbols in figure~\ref{diff}). Both methods of the estimation 
of diffusivity result in perfect agreement (figure~\ref{diff}). As it was seen from the Fourier-spectra 
at zero-frequency in figure~\ref{vacf_w}, the density dependence of the diffusion coefficient shows three 
regions: decay in the region of densities up to 0.4920 g/cm$^3$ due to reduction of the free 
volume in mainly molecular fluids, further increase of $D$ due to the rapid increase of the 
free volume because of the break-up of hydrogen molecules, while for densities higher than 0.7558 g/cm$^3$,
again there is observed a decay of $D$ due to the reduction of the free volume in mainly atomic fluid.
A relation between the free volume and diffusion coefficient was proposed by Bylander and Kleinman 
\cite{Byl92} and was used in \cite{Bry13} to explane the pressure dependence of diffusivity in 
liquid Rb across the structural transformation under pressure.

\begin{figure}[!t]
\centerline{
\includegraphics[width=0.65\textwidth]{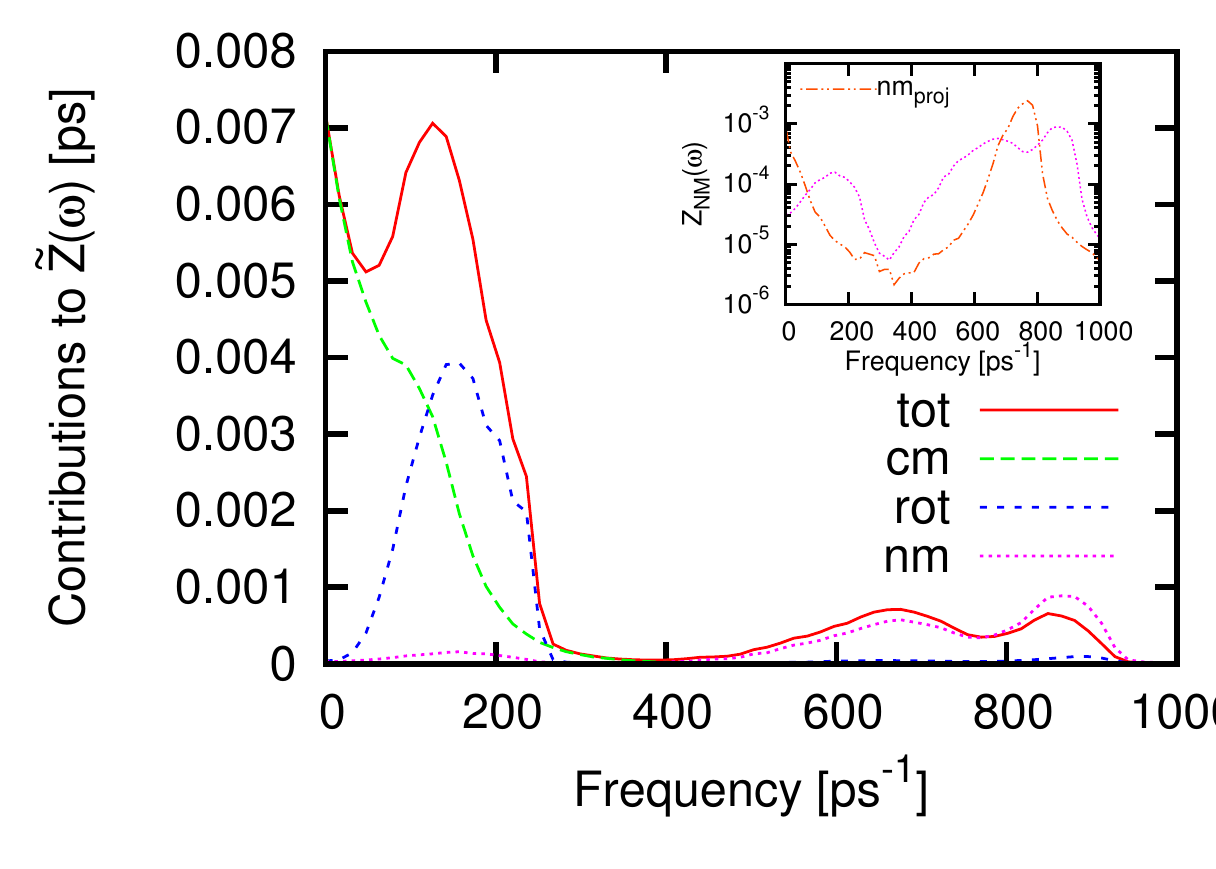}%
}
\caption{(Colour online) Contributions from the translational motion of the centers-of-mass (cm) and molecular 
rotation~(rot) around the centers-of-mass, and intramolecular normal modes (nm) to the Fourier 
spectrum of the velocity autocorrelation functions at 0.2847~g/cm$^3$ and $T=2500$~K. 
In the inset, the spectrum is shown for the case when the velocities in a H$_2$ molecule have 
been projected onto the molecular axis in order to remove the rotational component from 
the high-frequency peak.
} \label{contr}
\end{figure}

To assign  different peaks in the Fourier-spectrum of VACFs, we decomposed it into the 
contributions due 
to  translational motion of the centers-of-mass, rotational motion around the centers-of-mass, 
and the intra-molecular stretching vibrational normal mode. Such a decomposition is quite 
straightforward in the molecular frame adopting the vibrational normal mode to oscillations along
the molecular axis, while the rotation corresponds to the projection of the instantaneous proton 
velocity onto an orthogonal axis to the molecular one. In figure~\ref{contr}, we show this decomposition for
the case of the purely molecular fluid at the lowest density. The two high-frequency peaks are due to 
the intra-molecular stretching normal mode split into two as a result of the coupling to the 
rotational mode of the molecule. This can be proved by removing the rotational motion that results
in a single peak at 780~ps$^{-1}$, as is shown in the inset of figure~\ref{contr}. The low-frequency peak consists 
of contributions both from the rotation of the molecule and the translational motion, the latter 
extending to very low frequencies and being responsible for diffusivity. For higher densities, we
performed similar decompositions when the two neighbour protons were separated by distances not 
larger than 1.1~\AA.

\begin{figure}[!t]
\centerline{
\includegraphics[width=0.6\textwidth]{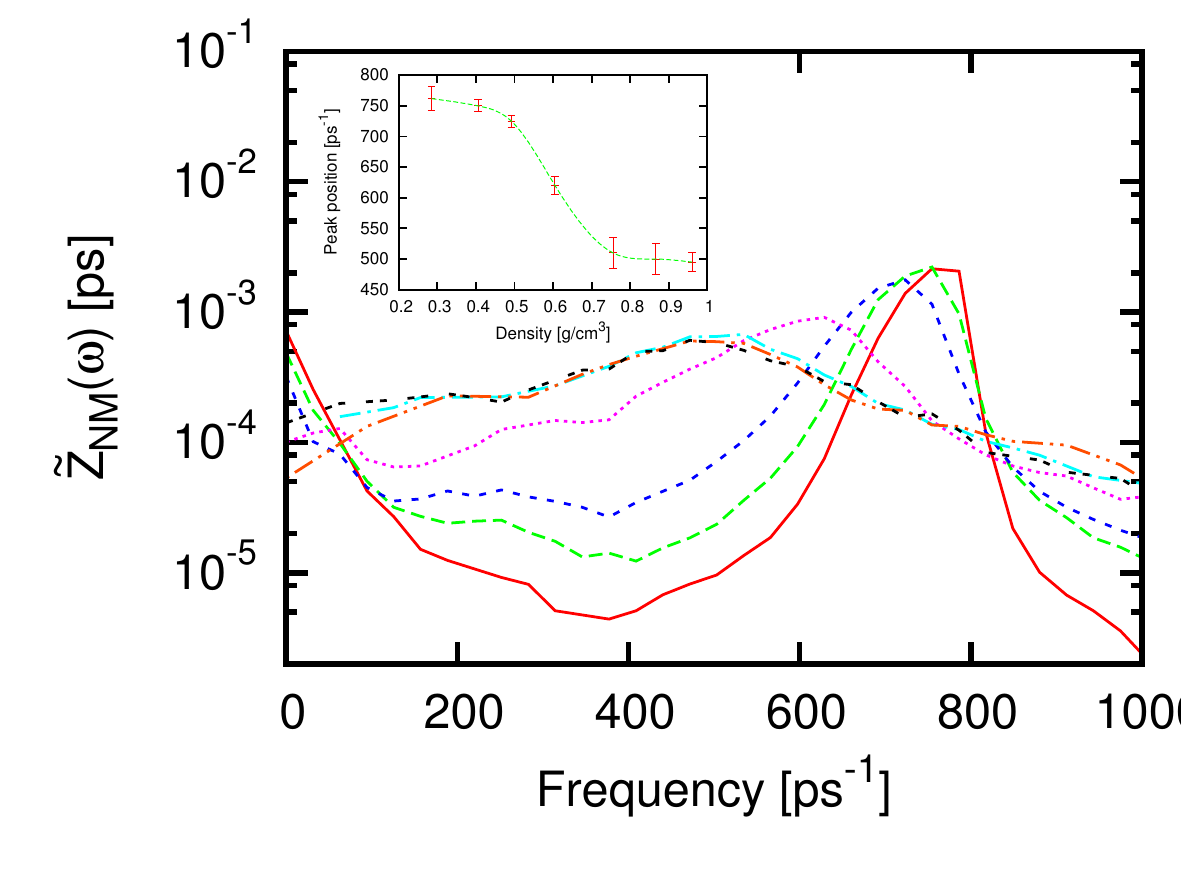}%
}
\caption{(Colour online) Evolution of the Fourier spectrum of normal modes with density.
The inset contains the pressure dependence of the maximum position of the normal mode distribution.
} \label{z_nm}
\end{figure}

In figure~\ref{z_nm}, we show the power spectrum of the intra-molecular stretching normal mode 
at different densities. An increase of density first leads  to a small reduction of the peak 
position, then to a rapid decrease in the frequency, and eventually to a transformation into a wide peak 
at 470~ps$^{-1}$, corresponding to the short-wavelength mode in the atomic fluid. Actually, these
three regimes are consistent with the non-monotonous change of the diffusion (in figure~\ref{diff}) across the
transformation from purely molecular to atomic hydrogen fluid. The maximum position of the normal 
mode distribution shows a decay from the state of molecular fluid with
well-defined intramolecular normal modes towards very smeared-out distributions of short-wavelength
modes in the atomic fluid with practically the same amplitude. This disappearance of the normal 
mode can be considered as some kind of the order parameter for the observed transition from molecular 
to atomic fluid in hydrogen under pressure.

\section{Conclusions}\label{section:conclusions}

{\it Ab initio} molecular dynamics simulations were performed for seven densities of the 
hydrogen fluid at 2500~K with the purpose to estimate the features in velocity autocorrelations 
across the transformation from purely molecular to atomic fluid. We observed a non-monotonous 
density dependence of diffusion coefficients across this transformation, which shows a region of 
densities with an increase of the diffusivity due to the break-up of hydrogen molecules. The velocity
autocorrelation functions reveal changes from the time dependence with high-frequency oscillations 
due to intramolecular normal modes to the typical monotonously decaying one for low-density atomic 
fluids, which, however, for 
the highest studied density (pressure) again shows a non-monotonous time decay. The
Fourier spectra of velocity autocorrelation functions were analyzed using a decomposition into 
contributions from translational and rotational motions as well as from normal 
intramolecular modes. The density dependence of contribution from 
 normal modes to vibrational spectra shows their gradual  disappearance and transformation of 
the corresponding high-frequency peak at high density into the one coming from the 
short-wavelength modes involving the most nearest neighbours in the purely atomic fluid.

\section*{Acknowledgements}
T.B. and A.P.S. were supported by the CNRS PISC-NASU project ``Ab initio simulations of
structural and dynamic features of complex and molecular fluids of geophysical interest''.
C.P. was supported by the Agence Nationale de la Recherche (ANR) France, under the program ``Accueil de Chercheurs de Haut Niveau 2015'' project: HyLightExtreme. Computer time was provided by the PRACE Project 2016143296, ISCRAB (IsB17\_MMCRHY) computer allocation at CINECA Italy, the high-performance computer resources from Grand Equipement National de Calcul Intensif (GENCI) Allocation 2018-A0030910282.
The calculations have been performed using the ab-initio total-energy
and molecular dynamics program VASP (Vienna ab-initio simulation package)
developed at the Institute f\"ur Materialphysik of the Universit\"at Wien
\cite{Kre93_1,Kre93_2,Kre96,Kre96b}.


%
\ukrainianpart

\title{Автокореляції швидкостей в області індукованого тиском перетворення молекулярний--атомний водневий флюїд}
\author{Дж. Руокко\refaddr{label1,label2}, Т. Брик\refaddr{label3,label4}, К. Пьєрлеоні\refaddr{label5,label6}, А.П. Сейтсонен\refaddr{label7}}
\addresses{
\addr{label1}
Центр нанонауки про життя @Сап'єнца, Італійський Інститут Технологій, \\вул. Королеви Елени, 295, I-00161, Рим, Італія
\addr{label2}
Фізичний факультет, Університет Риму Сап'єнца, I-00185, Рим, Італія
\addr{label3} Інститут фізики конденсованих систем НАН України, вул. Свєнціцького, 1,
79011 Львів, Україна
\addr{label4} Інститут прикладної математики та фундаментальних наук,
Національний університет ``Львівська Політехніка'', 79013 Львів, Україна
\addr{label5} Університет Париж-Саклє, Центр атомної енергiї,  Будинок симуляцiй, F-91191, Жiф-сюр-Iвет, Францiя
\addr{label6}
Факультет фізичних та хімічних наук, Університет Л'Аквіла, вул. Ветойо, 10, I-67010 Л'Аквіла, Італія
\addr{label7}
Хімічний факультет, Вища нормальна школа, вул. Льомон, 24, F-75005 Париж, Франція 
}

\makeukrtitle

\begin{abstract}
Немонотонні зміни в автокореляційних функціях швидкостей в області індукованого тиском перетворення молекулярний--атомний водневий флюїд
при температурі 2500~K досліджуються моделюванням методом першопринципної молекулярної динаміки. Ми повідомляємо поведінку коефіцієнтів 
дифузії в широкій області зміни густини: від чисто молекулярного флюїду аж до металічної фази атомного флюїду. Виконано аналіз внесків 
до автокореляційних функцій швидкостей від руху молекулярних центрів мас, обертових та внутрімолекулярних вібраційних мод, та обговорюється
кросовер у вібраційній густині інтрамолекулярних мод в області переходу.
	
\keywords перетворення молекулярний--атомний флюїд, водневий флюїд, автокореляційні функції швидкостей, першопринципна молекулярна динаміка
\end{abstract}


\begin{thebibliography}{12}
%
\bibitem{Mat09} Matsuoka T., Shimizu K., Nature, 2009, {\bf 458}, 186, \doi{10.1038/nature07827}. 
%
\bibitem{Ma09} Ma Y., Eremets M., Oganov A.R., Xie Y., Trojan I., Medvedev S.,
               Lyakhov A.O., Valle M., Prakapenka V., Nature, 2009, {\bf 458}, 182,
               \doi{10.1038/nature07786}. 
%
\bibitem{Bel17} Belonoshko A.B., Lukinov T., Fu J., Zhao J., Davis S., Simak S.I.,
               Nat. Geosci., 2017, {\bf 10}, 312, \doi{10.1038/ngeo2892}.
%
\bibitem{Bel19} Belonoshko A.B., Fu J., Bryk T., Simak S.I., Mattesini M.,  Nat. Commun., 2019,
               {\bf 10}, 2483,\\ \doi{10.1038/s41467-019-10346-2}.
%
\bibitem{Rat07} Raty J.-Y., Schwegler E., Bonev S.A., Nature, 2007, {\bf 449}, 448, 
                \doi{10.1038/nature06123}.
%
\bibitem{Tam08} Tamblyn I., Raty J.-Y., Bonev S.A., Phys. Rev. Lett., 2008, {\bf 101}, 075703, 
                \doi{10.1103/PhysRevLett.101.075703}. 
%
\bibitem{Bry13} Bryk T., De~Panfilis S., Gorelli F.A., Gregoryanz E., Krisch M.,
                Ruocco G., Santoro M., Scopigno T., Seitsonen~A.P.,
                Phys. Rev. Lett., 2013, {\bf 111}, 077801, \doi{10.1103/PhysRevLett.111.077801}. 
%
\bibitem{Ber}  Berne B.J., Pecora R., Dynamic Light Scattering, John Wiley, New York, 1976.
%
\bibitem{McM12} McMahon J.M., Morales M.A., Pierleoni C., Ceperley D.M., Rev. Mod. Phys.,
                2012, {\bf 84}, 1607,\\ \doi{10.1103/RevModPhys.84.1607}.
%
\bibitem{Bry20} Bryk T., Pierleoni C., Ruocco G., Seitsonen A.P., J. Mol. Liq., 2020, 113274, \doi{10.1016/j.molliq.2020.113274}.
%
\bibitem{Pie18} Pierleoni C., Holzmann M., Ceperley D.M., Contrib.  Plasma Phys., 2018, 
                \textbf{58}, 99, \doi{10.1002/ctpp.201700184}.
%
\bibitem{Duf94} Duffy T.S., Vos W.L., Zha C.-S., Hemley R.J., Mao H.-K., 
                Science, 1994, {\bf 263}, 1590,\\ \doi{10.1126/science.263.5153.1590}.
%
\bibitem{Ala95} Alavi A., Parrinello M., Frenkel D., Science, 1995, {\bf 269}, 1252, 
                \doi{10.1126/science.7652571}.
%
\bibitem{Mar} March N.H., Tosi M.P., Coulomb Liquids, Academic Press, London, 1984.
%
\bibitem{Bry04} Bryk T., Mryglod I., J. Phys.: Condens. Matter, 2004, {\bf 16}, L463, 
                \doi{10.1088/0953-8984/16/41/L06}.
\bibitem{Blo94} Bl\"ochl P.E., Phys. Rev. B, 1994, {\bf 50}, 17953, \doi{10.1103/PhysRevB.50.17953}.
\bibitem{Kre99} Kresse G., Joubert D., Phys. Rev. B, 1999, {\bf 59}, 1758, 
                \doi{10.1103/PhysRevB.59.1758}.
\bibitem{Per96} Perdew J.P., Burke K., Ernzerhof M., Phys. Rev. Lett., 1996, {\bf 77}, 3865,
                \doi{10.1103/PhysRevLett.77.3865}.
%
\bibitem{Han}   Hansen~J.-P., McDonald~I.R., Theory of Simple Liquids, Academic, London, 1986.
\bibitem{Boo} Boon J.-P., Yip S., Molecular Hydrodynamics, McGraw-Hill, New-York, 1980.
%
\bibitem{Byl92} Bylander D.M., Kleinman L., Phys. Rev. B, 1992, {\bf 45}, 9663, 
                \doi{10.1103/physrevb.45.9663}.
%
\bibitem{Kre93_1} Kresse G.,  Hafner J., Phys. Rev. B, 1993, {\bf 47}, 558, 
               \doi{10.1103/PhysRevB.47.558}.
\bibitem{Kre93_2} Kresse G.,  Hafner J., Phys. Rev. B, 1994, {\bf 49}, 14251,  
               \doi{10.1103/PhysRevB.49.14251}.
%
\bibitem{Kre96} Kresse G., Furthm\"uller J., Comput. Mater. Sci., 1996, {\bf 6}, 15, 
               \doi{10.1016/0927-0256(96)00008-0}.
%
\bibitem{Kre96b} Kresse G., Furthm\"uller J., Phys. Rev. B, 1996, {\bf 54}, 11169, 
                 \doi{10.1103/PhysRevB.54.11169}. 
\end{thebibliography}
\end{document}